\documentclass[aps,pra,10pt,
               preprintnumbers,numbers,sort&compress,nofootinbib,
               twocolumn,
               showpacs  
               ]{revtex4}
\usepackage{graphicx}
\usepackage{amsmath,amssymb,bm}

\newcommand{\ifprocessed}[2]{#2} 

\ifprocessed{\newcommand{\psfrag}[2]{}}{\usepackage{psfrag}}
\providecommand{\psfrag}[2]{}

\newcommand{\vect}[1]{\mathbf{#1}}

\newcommand{\abs}[1]{\lvert{#1}\rvert}
\newcommand{\op}[1]{\mathbf{\hat{#1}}}
\newcommand{\exclude}[1]{}
\renewcommand{\d}[0]{\mathrm{d}}

\newcommand{\pdiff}[3][{}]{\frac{\partial^{#1}{#2}}{\partial{#3}^{#1}}}

\newcommand{\g}{g}
\newcommand{\h}{h}
\newcommand{\x}{x}
\newcommand{\y}{y}
\newcommand{\Y}{Y}
\providecommand{\a}{}
\providecommand{\b}{}
\renewcommand{\a}{\ensuremath{a}}
\renewcommand{\b}{\ensuremath{b}}

\preprint{NT@UW-06-12}
\begin{document}

\title{Zero Temperature Thermodynamics of Asymmetric Fermi Gases
at Unitarity}
\author{Aurel Bulgac}
\author{Michael McNeil Forbes}
\affiliation{Department of Physics, University of Washington,
Seattle, WA 98195-1560}
\date{27 March 2007}
\begin{abstract}
  The equation of state of a dilute two-component asymmetric Fermi gas
  at unitarity is subject to strong constraints, which affect the
  spatial density profiles in atomic traps.  These constraints require
  the existence of at least one non-trivial partially polarized
  (asymmetric) phase.  We determine the relation between the structure
  of the spatial density profiles and the $T=0$ equation of state,
  based on the most accurate theoretical predictions available.  We
  also show how the equation of state can be determined from
  experimental observations.
\end{abstract}
\maketitle


We consider the $T=0$ thermodynamics of a dilute asymmetric Fermi gas
comprising two species of equal mass with $s$-wave interactions at
unitarity.  This has been recently realized in $^6$Li
experiments~\cite{ZSSK:2005,PLKLH:2005,ZSSK:2006,comments}.
We shall discuss the phase structure in the microcanonical and
grand-canonical ensembles, and its manifestation in cold atomic traps
using the local density approximation (LDA).\@  The theoretical
treatment of the grand-canonical
ensemble is much simpler than the microcanonical ensemble as it consists
of only pure phases.  We discuss here the most general model-independent
equation of state satisfying known constraints.  For model-dependent
analyses see~\cite{meanfield,Chevy:2006a,HS:2006}.

\begin{figure}[tbp]
  \centering
  \psfrag{Fully Paired (SF)}{Fully Paired (SF)}
  \psfrag{Vac}{Vac}
  \psfrag{Partially Polarized (PPa)}{Partially Polarized (PP$_{\a}$)}
  \psfrag{PPb}{PP$_{\b}$}
  \psfrag{Nb}{N$_{\b}$}
  \psfrag{Na}{N$_{\a}$}
  \psfrag{mua}{$\mu_{\a}$}
  \psfrag{mub}{$\mu_{\b}$}
  {\footnotesize
    \psfrag{0}{0}
    \ifprocessed{
      \includegraphics[width=\columnwidth]{figs_phase_diagram.eps}
    }{
      \includegraphics[width=\columnwidth]{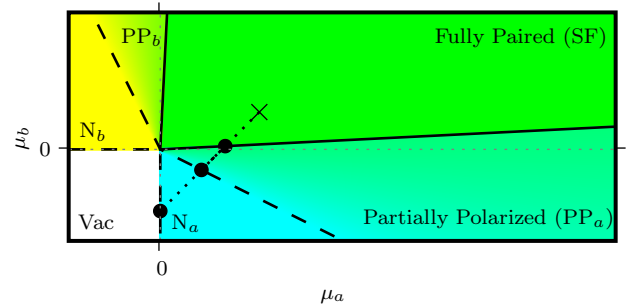}
    }
  }
  \caption{\label{fig:phase_diagram} Grand-canonical phase diagram
    of a two-component Fermi gas at unitarity and $T=0$.  Various
    phases are separated by phase transitions along the straight lines 
    extending from the origin with constant slopes $\y_{\x}$. 
    The dotted line follows the sequence of phases in a sample trap.}
\end{figure}

\paragraph{Phase structure:}
We show the main defining features of a grand-canonical phase diagram
in Fig.~\ref{fig:phase_diagram}.  The two species are labelled ``\a''
and ``\b''.  The symmetry $\a\leftrightarrow\b$ allows us to consider
only the region below the $\mu_{\a}=\mu_{b}$ line where the locally
averaged number densities and chemical potentials satisfy
$n_{\b}\leqslant n_{\a}$ and $\mu_{\b} \leqslant \mu_{\a}$
respectively.  The asymmetry of the system may thus be characterized
by the dimensionless ratios:
\begin{align}
  \x &= n_{\b}/n_{\a} \leqslant 1, &
  \y &= \mu_{\b}/\mu_{\a} \leqslant 1.
  \label{eq:1}
\end{align}
Note that only $\x$ measures a physical asymmetry.  There are four
distinct regions: Vac---the vacuum, N$_{\a}$---the fully polarized phases ($x\!=\!0$)
comprising only species ``\a'', PP$_{\a}$---partially
polarized phase(s) ($0\!<\!x\!<\!1$), and SF---the
fully paired symmetric superfluid phase ($x\!=\!1)$.

We shall discuss only two phase transitions: one between the fully
polarized phase N$_{\a}$ (where $\x=0$) and a partially polarized
phase PP$_{\a}$ ($0<n_{\b}<n_{\a}$), and another between a (possibly
different) partially polarized phase and the symmetric fully paired
phase SF (where $\x=1$).  At unitarity, phase transitions occur along
rays characterized solely by their slope $\y_{\x}$.  The two
transitions we shall discuss are thus described by the two universal
parameters $\y_{0}$ and $\y_{1}$, which naturally satisfy $\y_{0}
\leqslant\y_{1}$.  A major point of this paper is to place an upper
bound $\Y_0$ on $\y_0$ ($\y_0 \leqslant \Y_0$), a lower bound
$\Y_1$ on $\y_1$ ($\Y_1\leqslant\y_1$), and to show $\Y_0<\Y_1$,
which implies that the inequality $\y_{0}<\y_{1}$ is strict.  This
directly implies the existence and stability of one or more partially
polarized phase(s) PP$_{\a}$.  Possible phases in the region PP$_{\a}$
include LOFF states~\cite{loff}, states with deformed Fermi
surfaces~\cite{Muther:2002mc}, and $p$-wave superfluid
states~\cite{Bulgac:2006gh}.  If several of these states exists and
are stable, the corresponding phase transitions will be characterized
by additional universal parameters $\y_{\x}$.

\paragraph{Functional forms of thermodynamic potentials:}
At unitarity, the energy density $\mathcal{E}(n_{\a},n_{\b})$ and the
pressure $\mathcal{P}(\mu_{\a},\mu_{\b})$ have the following form:
\begin{subequations}
  \label{eq:EP}
  \begin{align}
    \mathcal{E}(n_{\a},n_{\b}) 
        &= \tfrac{3}{5}\alpha\left[n_{\a}\g(\x)\right]^{5/3}, &
    \alpha &= \frac{(6\pi^2)^{2/3}\hbar^2}{2m}, \label{eq:E} \\
    \mathcal{P}(\mu_{\a},\mu_{\b}) 
        &= \tfrac{2}{5}\beta\left[\mu_{\a}\h(\y)\right]^{5/2}, &
    \beta &= \frac{1}{6\pi^2}\left[\frac{2m}{\hbar^2}\right]^{3/2}. \label{eq:P}
  \end{align}
\end{subequations}
Note that $\g(\x) = f^{3/5}(\x)$, where $f(x)$ was introduced
in~\cite{Cohen:2005ea}: The use of $\g(\x)$ rather than $f(x)$
significantly simplifies the formalism~\cite{appendix}.  The $T=0$
thermodynamic properties of the system are completely determined by
the functional form of $\g(\x)$ or $\h(\y)$.  The number densities and
chemical potentials are simply $n_{a,b} = \partial\mathcal{P}/\partial
\mu_{a,b}$ and $\mu_{a,b} = \partial\mathcal{E}/\partial n_{a,b}$
respectively.  As we show here, the functions $\h(\y)$ and $\g(\x)$
are tightly constrained by current Monte Carlo simulations, analytic
calculations, and experimental results.  The energy density and
pressure are related via the Legendre transform~\cite{appendix}:
\begin{equation}
  \label{eq:Legendre}
  \mathcal{P}(\mu_{\a},\mu_{\b}) = \mu_{\a} n_{\a}+\mu_{\b} n_{\b}-\mathcal{E}(n_{\a},n_{\b})
  =\tfrac{2}{3}\mathcal{E}(n_{\a},n_{\b}).
\end{equation}


\paragraph{Physical constraints:}
The thermodynamics of three phases are known.  The vacuum has
vanishing pressure $\mathcal{P}_{\text{Vac}}=0$, the fully polarized
phase N$_{\a}$ is a free Fermi gas,
\begin{equation}
  \label{eq:PFG}
  \mathcal{P}_{\text{FG}}(\mu_{\a}) = \tfrac{2}{5}\beta \mu_{\a}^{5/2},
\end{equation}
and the pressure of the fully paired phase SF is symmetric in
$\mu_{\a}$ and $\mu_{\b}$, and is described by a single parameter $\xi$,
\begin{equation}
  \label{eq:PSF}
  \mathcal{P}_{\text{SF}}(\mu_{+}) = 
  \frac{4}{5} \frac{\beta}{\xi^{3/2}}\mu_{+}^{5/2} \quad \text{where} \quad
  \mu_{\pm} = \frac{\mu_{\a}\pm\mu_{\b}}{2}.
\end{equation}
These provide the limiting forms for $\h(\y)$ and limiting values of
$\g(\x)$, [see Eqs. (\ref{eq:h_lim}) and (\ref{eq:g01}) below].

The phase transition at $\y\!=\!\y_{0}$ defines the border of the
region where $\mu_{\b} = \y_{0}\mu_{\a}$ is tuned to
keep species ``\b'' out of the system.  Since the interspecies
interaction is attractive, the critical $\mu_{\b}$ must be
negative $\y_{0} < 0$.  We will provide an upper
bound $\Y_{0}$ ($\y_{0} \leqslant \Y_{0}$).

Note that $\mathcal{P}_{\text{SF}}(\mu_{+})$ depends only on the
average chemical potential $\mu_{+}$.  This insensitivity to the
chemical potential difference $\mu_{-}$ is due to the existence of an
energy gap $\Delta$ in the spectrum.  The phase transition at
$\y\!=\!\y_{1}$ marks the line where $\mu_{-}$ becomes large enough to
break the superfluid pairs.  In the phase SF, $\mu_{-}$ is constrained
by the size of the physical gap $\mu_{-} \leqslant
\Delta$~\cite{Cohen:2005ea}.  This provides a lower bound
$\Y_{1}\leqslant \y_{1}$, see below and \cite{appendix}.  

(Recall from Eq.~(\ref{eq:1}) that we are only
considering regions where $0\leq\mu_{-}=\abs{\mu_{-}}$).

If no stable partially polarized phase exists, then the region
PP$_{\a}$ will vanish, being compressed into a single first-order
transition line where pressure equilibrium is established
$\mathcal{P}_{\text{FG}}(\mu_{\a})
=\mathcal{P}_{\text{SF}}(\mu_{+})$~\cite{Cohen:2005ea,Chevy:2006a}.
This would occur at $\y=\y_c= (2\xi)^{3/5}-1\equiv\y_{0}\equiv\y_{1}$,
and would imply that $\Y_{0}\equiv\Y_{1}$.  We argue below that $\Y_{0}$ is
strictly less than $\Y_{1}$, and therefore rule out this possibility.

Finally, thermodynamic stability requires that the pressure and energy
density are convex functions, which implies that $g(x)$ and $h(y)$ are
also convex~\cite{appendix}.  The constraints on $\h(\y)$ are
\begin{subequations}
  \label{eq:h}
  \begin{align}  
    \h(\y) &= \begin{cases}
      1 & \text{if } \y \leqslant \y_{0},\\
      (1+\y)(2\xi)^{-3/5} &\text{if } \y \in [\y_{1},1],
    \end{cases}\label{eq:h_lim}\\  
    h''(y) &\geqslant 0, \qquad\text{and}\label{eq:h_concave}\\
    \y_{0} &\leqslant \Y_{0}<\y_{c}<\Y_{1} \leqslant \y_{1}
    \leqslant 1. \label{eq:h_inequal}
  \end{align}
\end{subequations}
The corresponding constraints on $g(x)$ are
\begin{subequations}
  \label{eq:g}
  \begin{align}
    g(0) &= 1, &
    g(1) &= (2\xi)^{3/5},\label{eq:g01}\\
    g''(x) &\geqslant 0, &&\text{ and} \label{eq:g_concave}\\
    g'(0) &\leqslant \Y_{0},&
    g'(1) &\in \bigl[g(1)(1+\Y_{1}^{-1})^{-1},g(1)/2\bigr]. \label{eq:g'}
  \end{align}
\end{subequations}
Equation~(\ref{eq:h_lim}) follows directly from Eqs.~(\ref{eq:P}),
(\ref{eq:PFG}), and (\ref{eq:PSF}), Eq.~(\ref{eq:g01}) follows from
Eq.~(\ref{eq:E}), and the interval in Eq.~(\ref{eq:g'}) follows from
the properties of the Legendre transform and Eq. (\ref{eq:h_inequal})
\cite{appendix}.


\begin{figure}[tbp]
  \centering
  \psfrag{y0=y0}{$\y_{0}=\Y_{0}$}
  \psfrag{yc=y1}{$\y_{c}\approx\Y_{1}$}
  \psfrag{y1}{$\y_{1}$}
  \psfrag{y=mub/mua}{$\y=\mu_{\b}/\mu_{\a}$}
  \psfrag{hy}{$\h(\y)$}
  \psfrag{x=nb/na}{$\x=n_{\b}/n_{\a}$}
  \psfrag{gx}{$\g(\x)$}
  {\footnotesize
    \psfrag{0.8}{0.8}
    \psfrag{0.9}{0.9}
    \psfrag{1.0}{1.0}
    \psfrag{1.1}{1.1}
    \psfrag{1.2}{1.2}
    \psfrag{1.3}{1.3}
    \psfrag{-0.6}{-0.6}
    \psfrag{-0.4}{-0.4}
    \psfrag{-0.2}{-0.2}
    \psfrag{0.0}{0.0}
    \psfrag{0.2}{0.2}
    \psfrag{0.4}{0.4}
    \psfrag{0.6}{0.6}
    \psfrag{0.8}{0.8}
    \ifprocessed{
      \includegraphics[width=\columnwidth]{figs_gh.eps}
    }{
      \includegraphics[width=\columnwidth]{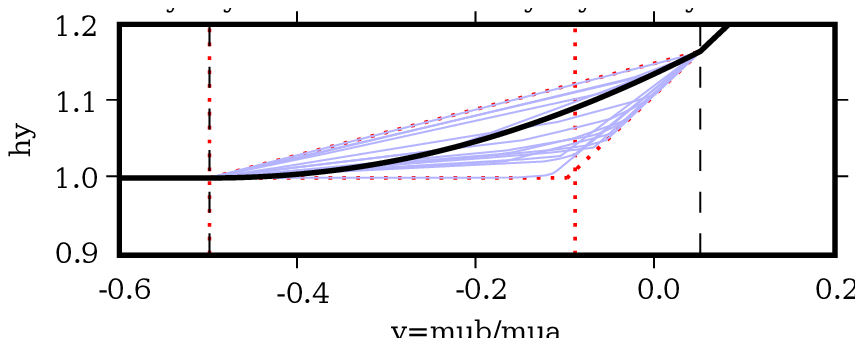}
      \includegraphics[width=\columnwidth]{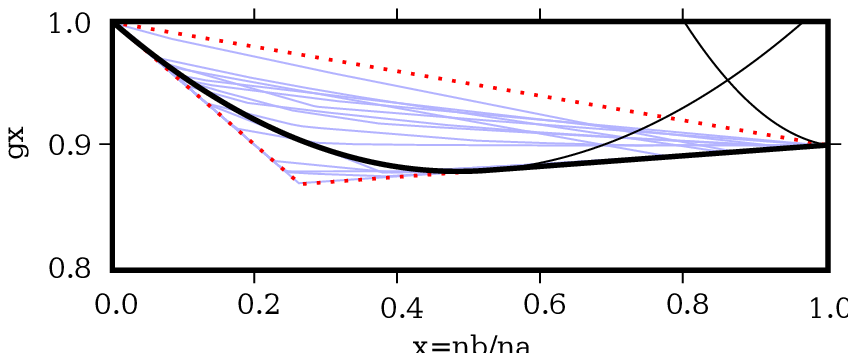}
    }
  }
  \caption{\label{fig:hg} Example of a function $\h(y)$ and the
    corresponding function $\g(\x)$ shown as thick lines.  Maxwell's
    construction for phase coexistence leads to a linear $\g(\x)$ for
    $\x\in(0.5,1.0)$, interpolating between the two pure phases shown
    with lighter lines.  This corresponds to the kink (first-order
    phase transition) at $\y=\y_{1}$ in $\h(\y)$.  Various other
    sample functions are lightly sketched within the allowed (dotted)
    triangular region.  }
\end{figure}
\paragraph{Parameters:}
For Fig.~\ref{fig:hg} we used
\begin{align*}
  \xi &= 0.42(1), &
  \Y_{0} &\approx -0.5,&
  \Y_{1} &= -0.09(3), &
  \y_{1} &= 0.05.
\end{align*}
We obtain estimates for $\Y_{1}$ and $\xi$ from Monte Carlo
data~\cite{Pandharipande,ABCG:2004,Carlson:2005kg}.  The latest
Monte Carlo estimates for the symmetric systems give $\xi =
0.42(1)$~\cite{ABCG:2004,Carlson:2005kg} and
$\Delta/\varepsilon_{\text{F}} = 0.504(24)$~\cite{Carlson:2005kg},
where $\varepsilon_{\text{F}}$ is the Fermi energy of the free gas
with the same density.  This gives $\y_{c} \approx -0.099(15)$, and
the constraint $\mu_{-}\!\leqslant\!\Delta$ gives $\Y_{1} =
(\xi-\Delta/\varepsilon_{\text{F}})/(\xi+\Delta/\varepsilon_{\text{F}})
= -0.09(3)$~\cite{Cohen:2005ea,appendix}.  Since within the
statistical errors $Y_1\approx y_c$, the possibility of an empty
PP$_{\a}$ region at unitarity cannot yet be ruled out by this value of
$\Y_{1}$, as was noted earlier by Cohen~\cite{Cohen:2005ea}.

We now estimate $\Y_{0}$.  Let $e_{0}$ be the energy required to add
one particle $b$ to a fully polarized gas of density $n_{\a}$.
Consider adding a large, but infinitesimal amount of $b$, $1\ll
N_{\b}\ll N_{\a}$. In the thermodynamic limit, the required energy per
particle will be the critical chemical potential $\mu_{\b} = \alpha
n_{\a}^{2/3} g'(0)$ defining the transition $\y_{0}$.  If the added
particles repel, the energy will be $N_{\b}e_{0}$, and
$\mu_{\b}\!=\!e_{0}$.  If they bind, the additional binding energy
must be included, giving $\mu_{\b} < e_{0}$.  In this way, $e_{0}$
provides a bound for $\mu_{\b}$ and
$\smash{g'(0)=\y_{0}\leqslant\Y_{0}=e_{0}/\bigl(\alpha
  n_{\a}^{2/3}\bigr)}$.

Consider adding a single \b{} fermion, with coordinate $\vect{r}_{0}$,
to a system of $N_{\a}$ \a{} fermions with coordinates $\vect{r}_{n}$.
Let $r_{nm} = \abs{\vect{r}_{n}-\vect{r}_{m}}$.  The wave function for
the \b{} fermion in the background of fixed \a{} sources is
\begin{equation}
  \phi(\vect{r}_{0};\{\vect{r}_{n}\}) = \sum_{n} A_{n}
  \frac{\exp{(-\kappa r_{0n})}}{r_{0n}}, \label{eq:phi0}
\end{equation}
and it satisfies the zero-range interaction boundary conditions if the
following $N_{\a}$ conditions are met:
\begin{equation}
  \left(-\kappa+\frac{1}{a}\right)A_{n} + \sum_{m\neq n}
  A_{m}\frac{\exp({-\kappa r_{nm}})}{r_{nm}} = 0. \label{eq:phi1}
\end{equation}
For uniform distributions where the lowest state has
constant $A_{m}=A$, approximating the sum as an integral gives
$\kappa-a^{-1} = 4\pi n_{\a}/\kappa^2$.  This continuum approximation
is not very accurate in the unitary limit, since $\kappa$ is
comparable to the inverse interparticle spacing and $\kappa^3/n_{a}
\approx 4\pi$.  To estimate corrections, the equations can be solved
for various lattice configurations.  We find that $\kappa$ deviates
from the continuum result by a factor of $0.84(3)$ for simple lattice
configurations and perturbations (see \cite{appendix} for details).
We now estimate the energy of the system using the wave function
\begin{equation}
  \Psi(\vect{r}_{0};\{\vect{r}_{n}\}) = 
  \Phi_{\text{SD}}(\{\vect{r}_{n}\})\phi(\vect{r}_{0};\{\vect{r}_{n}\}),
  \label{eq:wf}
\end{equation}
where $\Phi_{\text{SD}}$ is a Slater determinant for a free Fermi gas
and obtain $e_{0} \approx -\hbar^2\kappa^2/m$~\cite{appendix}:
\begin{equation}
  e_{0} \approx
  \begin{cases}
    4\pi \hbar^2n_{\a} a/m,
    & \text{if } a \rightarrow 0^{-},\\
    -0.71(5)\hbar^2 (4\pi n_{\a})^{2/3}/m ,
    & \text{if } a \rightarrow \pm\infty, \\
    -\hbar^2/(ma^2), 
    & \text{if } a \rightarrow 0^{+}.
  \end{cases}
\end{equation}
Note that this result interpolates between the correct leading order
BEC and BCS results.  This estimate assumes that the fluctuations of
the number density $n_{\a}(\vect{r})$ on a scale of the order
$1/\kappa$ affect $\kappa$ very little.  The result is consistent with
this assumption, as discussed in~\cite{appendix}.  The constraint at
unitarity is thus
\begin{equation}
  \label{eq:Y0}
  \Y_{0} \approx -0.54(4) <
  \y_{c}=-0.099(15).
\end{equation}
If $\Y_{0}$ is strictly less than $\y_{c}$, then convexity in $\g(\x)$
and $\h(\y)$ implies $\y_c<\Y_1$ (see Fig.~\ref{fig:hg}).


\paragraph{Trap profiles:}
For large systems with a slowly varying confining potential, gradient
terms may be neglected, and the LDA employed to determine the density
distribution by introducing spatially varying effective chemical
potentials:
\begin{equation}
  \label{eq:mu_r}
  \mu_{\a,\b}(\vect{R}) = \lambda_{\a,\b}-V(\vect{R}).
\end{equation}
Lagrange multipliers $\lambda_{\a,\b}$ fix the total particle numbers
$N_{\a}$ and $N_{\b}$.  The LDA may be inaccurate near phase boundaries where
the densities change rapidly.  The gradient terms will smear out these
transition regions and provide an additional surface tension
proportional to the local curvature~\cite{grad-lda}.
\begin{figure}[tbp]
  \centering
  \psfrag{Total Polarization |Na-Nb|/(Na+Nb)}
  {Total Polarization $\abs{N_{\a}-N_{\b}}/(N_{\a}+N_{\b})$}
  \psfrag{g}{$\gamma$}
  \psfrag{R0/Rvac, R1/Rvac}
  {$R_{0}/R_{\text{vac}}$, $R_{1}/R_{\text{vac}}$}
  {\footnotesize
    \psfrag{0.0}{0.0}
    \psfrag{0.2}{0.2}
    \psfrag{0.4}{0.4}
    \psfrag{0.6}{0.6}
    \psfrag{0.8}{0.8}
    \psfrag{1.0}{1.0}
    \ifprocessed{
      \includegraphics[width=\columnwidth]{figs_data.eps}
    }{
      \includegraphics[width=\columnwidth]{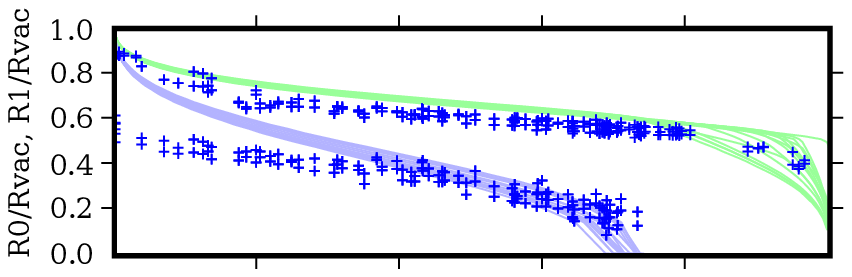}
      \includegraphics[width=\columnwidth]{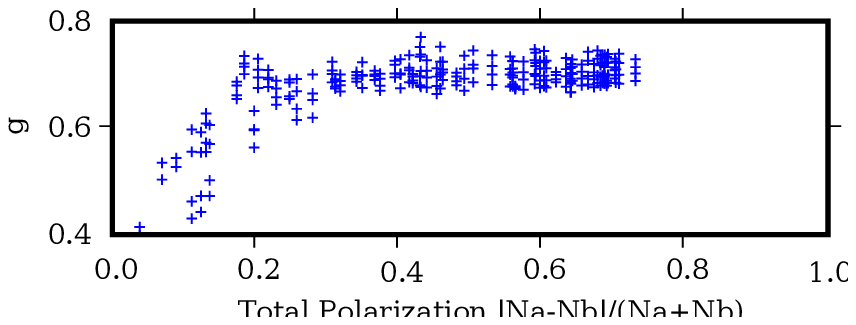}
    }
  }
  \caption{\label{fig:R} Measured transition radii
    from Ref. \cite{ZSSK:2006}.  The upper plot shows the normalized
    data (crosses) on top of data generated from several randomly
    generated functions $h(y)$.  The lower plot shows the parameter
    $\gamma$ defined in Rel. (\ref{eq:y0y1rel}).}
\end{figure}

In the LDA, the density profile may be constructed
from the local densities $n_{\a,\b}$ using Eq.~(\ref{eq:mu_r}) (explicit
formulae are provided in \cite{appendix}).  The dotted line in
Fig.~\ref{fig:phase_diagram} shows the sequence of phases contained in a sample
trap.  Since $2\mu_{-}=\lambda_{\a}-\lambda_{\b}$ is fixed, traps
contain the sequence of phases encountered along a $45^\circ$ line through such a diagram.  In
this example, the center of the cloud ($V(\vect{0})\!=\!0$) is in the
SF phase at the cross.  The phase transitions will occur for
$\y\!=\!\y_{1}$, $\y\!=\!\y_{0}$, and $\y=-\infty$ for
$V(\vect{R}_1)\!=\!V_{1}$, $V(\vect{R}_{0})\!=\!V_{0}$, and
$V(\vect{R}_{\text{vac}})\!=\!V_{\text{vac}}$ respectively, with
$R_{0}<R_{1}<R_{\text{vac}}$.  As noted above, additional phase
transitions may exist between $R_{0}$ and $R_{1}$.


\paragraph{Experiments:}
Accurate measurements of the density profiles would allow for a
complete extraction of $\h(\y)$ and $\g(\x)$.  For example, using $\x
= n_{\b}(\vect{R})/n_{\a}(\vect{R})$ and the expressions for
$\mu_{\a,\b}$, we have $\g^{2/3}(\x)\g'(\x) =
[\lambda_{\b}-V(\vect{R})]/[\alpha n_{\a}^{2/3}(\vect{R})]$ from which
$g(x)$ may be extracted using the boundary condition
$\g(0)=1$~\cite{appendix}.

For harmonic traps, the locations of the main phase transitions,
$R_{\text{vac}} \propto \sqrt{V_{\text{vac}}}$, $R_{0} \propto
\sqrt{V_{0}}$, and $R_{1} \propto \sqrt{V_{1}}$, are completely
determined by the Lagrange multipliers $\lambda_{\a,\b}$, and the
universal numbers $\y_{0}$ and $\y_{1}$.  Within the LDA, we obtain
the following model independent relationship, to be compared with the
recent MIT data~\cite{ZSSK:2006} (see Fig.~\ref{fig:R}):
\begin{equation}
  \label{eq:y0y1rel}
  \gamma = \frac{1-\y_{1}}{1-\y_{0}} = \frac
  {R^2_{\text{vac}}-R^2_{0}}
  {R^2_{\text{vac}}-R^2_{1}}
  \approx 0.70(5).
\end{equation}
To extract more information, one must consider a specific functional
form for $\h(\y)$ and $\g(\x)$.  We have analyzed a large sample of
allowed functions $\h(\y)$ and $\g(\x)$, a few of which are
sketched in Figs.~\ref{fig:hg} and~\ref{fig:R}.  We find that the
total polarization $P=(N_{\a}-N_{\b})/(N_{\a}+N_{\b})$ is quite
insensitive to the functional form.  However, the critical
polarization $P_{c}$---where the innermost phase transition approaches
the center of the trap $R_{1}=0$---is quite sensitive to $\y_{1}$.
The MIT experiments~\cite{ZSSK:2005,ZSSK:2006} measure $P_{c} =
0.70(5)$.  If $\y_{1}=0$, then one obtains $P_{c} > 0.80\ldots 0.85$.
However, if one considers $\y_{1} \approx 0.05$ ($g'(1) \approx
0.04$), then values of $P_{c}\approx 0.7$ and smaller emerge,
compatible with those measured in~\cite{ZSSK:2005,ZSSK:2006}.
Using Eq.~(\ref{eq:y0y1rel}), this gives a value of $\y_{0} = g'(0)
\approx -0.4$.  Our estimate for $Y_0\!\approx\!-0.54(4)$ is
consistent with this extracted experimental value within existing
uncertainties.

Within the Eagles-Leggett extension of the BCS model~\cite{eagles},
one obtains the values $\y_0=0$, $\y_c= 0.105$, and $\y_1= 0.107$
(see~\cite{HS:2006,appendix}), which would correspond to a parameter
$\gamma = 0.893$, as opposed to the $\gamma =0.70(5)$ extracted from
experiment.  At the same time, the spatial layer for the PP$_a$ region
would be very thin, namely
$(R_{0}^2-R_{1}^2)/(R_{\text{vac}}^2-R_{1}^2) = 1-\gamma = 0.107$,
compared with our estimate $1-\gamma \approx 0.30(5)$.

Our analysis is strictly valid only at $T=0$. The deviations
in Fig.~\ref{fig:R} are most likely finite temperature effects. The
regions of the phase diagram most sensitive to $T\neq 0$ are those
with small $\mu$, thus, the transition radii in traps with small
asymmetry will be most affected.  For large polarizations, the
temperature should not affect the SF phase which is gapped, but will
affect phases with zero or small gap excitations.  This could alter
the values of $\y_{1}$ slightly and $\y_{0}$ significantly.  We thus
caution against taking the extracted numbers in this section too
seriously until a similar finite temperature analysis is presented.

Since $\Y_{1} < 0$ (as determined from the value of the pairing gap),
a positive $\y_{1}$ suggests a first-order phase transition out of the
SF phase for a critical $\mu_{-} < \Delta$.  This conclusion is also
consistent with the form of the quasiparticle spectrum computed
in~\cite{Carlson:2005kg}, which has a minimum at a finite momentum.

In conclusion, we have shown that thermodynamic constraints, accurate
Monte Carlo simulations, analytic estimates, and experimental data
place tight constraints on the equation of state of the asymmetric
$T=0$ unitary gas.  These constraints imply that there exists a region
where \emph{one or more nontrivial partially polarized phases exist}.
These phases likely exhibit very interesting microscopic physics.  In
particular, any ungapped polarized phase is unstable towards the
formation of a state with two symbiotic superfluids at
$T=0$~\cite{Bulgac:2006gh}.  The tight constraints on the forms of
$\g(\x)$ and $\h(\y)$ we present will help locate these novel phases.

The authors thank A.~Schwenk and M.~W.~Zwierlein for comments,
D.~T.~Son for many useful discussions, M.~W.~Zwierlein \textit{et
  al.} for providing the data in Fig.~\ref{fig:R}, and the US
Department of Energy for support under Grant No. DE-FG02-97ER41014.

\textit{Notes added:} Chevy~\cite{Chevy:2006b} independently arrived at
similar conclusions.  The latest MIT analysis \cite{SZSSK:2006} agrees
with our conclusion that the SF phase occupies the center of the trap
and shows that the LDA is applicable.  In a recent variational Monte
Carlo study, Lobo {\it et al.}~\cite{LRGS:2006} agree with our lower bound,
obtaining $Y_0=-0.58(1) < y_c=-0.099(15)$, and concluded that the
transition at $y_1$ is first-order.  This is consistent with our
results: their function $f(x)$ is very similar to our $g^{5/3}(x)$
(see~\cite{appendix}).



\appendix
\section{Additional Details}
In this appendix, we present some additional calculational details.
We discuss the properties of the Legendre transformation,
thermodynamic stability and convexity of thermodynamic potentials, and
the Maxwell construction for mixed phases.  We also provide details of
the calculations whose results are presented in the main body.
\subsection{Properties of the Legendre Transformation}
Many of the thermodynamic constraints follow directly from properties
of the Legendre transformation~(\ref{eq:Legendre}).  We explain some
of these properties here, because there has been some confusion in the
literature.  

The Legendre transform relates tangents in one ensemble to coordinates
in the other.  For example, a linear segment of $g(\x)$ over a finite
interval in $\x$ will be mapped into a single point $\y$,
where $\h(\y)$ has a kink.  Straight segments
of $\g(\x)$ arise from the Maxwell construction and indicate a phase
coexistence (mixed phase).  For this reason, the grand-canonical phase
diagram is much simpler than other ensembles as it consists solely
of pure phases.

First we present some relations.  Starting with the
definitions of the thermodynamic potential densities
$\mathcal{E}(n_{\a},n_{\b})$ and $-\mathcal{P}(\mu_{a},\mu_{b})$
defined in~(\ref{eq:EP}), we differentiate to find the densities
\begin{subequations}
  \label{eq:densities}
  \begin{align}
     n_{\a} &= \pdiff{\mathcal{P}}{\mu_{\a}} = \beta [\mu_{\a}\h(\y)]^{3/2}
    [\h(\y)-\y \h'(\y)],\\
    n_{\b} &= \pdiff{\mathcal{P}}{\mu_{\b}} = \beta [\mu_{\a}\h(\y)]^{3/2}
    \h'(\y),
  \end{align}
\end{subequations}
and the chemical potentials
\begin{subequations}
  \label{eq:mu}
  \begin{align}
    \mu_{\a} &= \pdiff{\mathcal{E}}{n_{\a}} = 
    \alpha [n_{\a}g(x)]^{2/3}[g(x)-xg'(x)],\\
    \mu_{\b} &= \pdiff{\mathcal{E}}{n_{\b}} = 
    \alpha [n_{\a}g(x)]^{2/3}g'(x).\label{eq:mub}
  \end{align}
\end{subequations}
From these relations and the definitions of the asymmetry parameters
$\x=n_{\b}/n_{\a}$ and $\y=\mu_{\b}/\mu_{\a}$, we obtain the following
dictionary.  These relations may be used to express $g(x)$ and $x$ in
terms of $h(y)$ and $y$ or vice versa:
\begin{subequations}
  \label{eq:dictionary}
  \begin{align}
    \y &= \frac{\g'(\x)}{\g(\x)-\x\g'(\x)}, &
    \h(\y) &= \frac{1}{\g(\x)-\x\g'(\x)}, \\
    \x &=\frac{\h'(\y)}{\h(\y)-\y\h'(\y)}, &
    \g(\x) &= \frac{1}{\h(\y)-\y\h'(\y)}.
  \end{align}
\end{subequations}
The important geometric property of the Legendre transformation is
that it maps tangents in one space to points in the other.  Consider
the grand-canonical ensemble described by the function
$\mathcal{P}(\mu_{\a},\mu_{\b})$.\footnote{Strictly speaking the grand-canonical
potential is $-V\mathcal{P}(\mu_{\a},\mu_{\b})$, notice the minus sign, 
where $V$ is the volume.} The tangents
$(\partial_{\mu_{\a}}\mathcal{P},\partial_{\mu_{\b}}\mathcal{P})=(n_{\a},n_{\b})$
to this function at a point $(\mu_{a},\mu_{\b})$ in the
grand-canonical ensemble map directly to the coordinate
$(n_{\a},n_{\b})$ in the microcanonical ensemble.  Note that the
Legendre transformation is symmetric: tangents in the microcanonical
ensemble map back to points in the grand-canonical ensemble.
\exclude{
restart;
with(VectorCalculus):with(linalg):

Ef:=3/5*(a^(5/3))*f(b/a);
HEf:=Hessian(Ef,[a,b]):
ssf:={f(b/a)=f,D(f)(b/a)=df,D(D(f))(b/a)=ddf,b=a*x}:
H0:=simplify(subs(ssf,HEf)*5/a^3);

Eg:=3/5*(a*g(b/a))^(5/3);
HEg:=Hessian(Eg,[a,b]):
ssg:={g(b/a)=g,D(g)(b/a)=dg,D(D(g))(b/a)=ddg,b=a*x}:
H1:=simplify(subs(ssg,HEg)*6*(a*g)^(1/3));

P:=2/5*(a*h(b/a))^(5/2);
HP:=Hessian(P,[a,b]):
ssh:={h(b/a)=h,D(h)(b/a)=dh,D(D(h))(b/a)=ddh,b=a*y}:
H2:=simplify(subs(ssh,HP)*2/(a*h)^(1/2));
}

\subsection{Thermodynamic Stability: The Second Law}

The thermodynamic potentials follow from a strict minimization
procedure over all possible states.  If this is properly carried out,
the potentials will be convex functions of their arguments.  This
convexity is the geometric manifestation of the second law.  Locally,
the requirement is that the Hessian of the potentials---the matrix of
second partial derivatives---be positive semi-definite.  One can
easily show that this requirement also implies that both density and
concentration sound modes are stable with corresponding real sound
velocities.  We shall evaluate now the Hessian's for the thermodynamic
potential densities in the case of a two species at unitarity and
$T=0$. We thus find that
\begin{align*}
  H(\mathcal{P}_h) &\propto
  \begin{bmatrix}
    3(h-yh')2+2y2 h h'' & 3 (hh' -yh'^2)-2yhh''\\
    3 (hh'-yh'^2)-2yhh'' & 3h'^2+2hh''
  \end{bmatrix},\\
  H(\mathcal{E}_g) &\propto
  \begin{bmatrix}
    4(g-xg')2+6x2 g g'' & 4 (gg' -xg'^2)-6xgg''\\
    4 (gg' -xg'^2)-6xgg'' & 4g'^2+6gg''
  \end{bmatrix},\\
  H(\mathcal{E}_f) &\propto
  \begin{bmatrix}
    10f-12xf'+9x^2f'' & 6f'-9xf''\\
    6f'-9xf'' & 9f''
  \end{bmatrix}.
\end{align*}
The corresponding determinants are:
\begin{align}
  \det[H(\mathcal{P}_h)] &\propto h^3 h''\\
  \det[H(\mathcal{E}_g)] &\propto g^3g''\\
  \det[H(\mathcal{E}_f)] &\propto 5 f f'' - 2 f'^2.
\end{align}
From these it is easy to see that the Hessians of $\mathcal{P}$ and
$\mathcal{E}$ are positive definite if $h''\geqslant 0$ and
$h\geqslant 0$, or $g''\geqslant 0$ and $g\geqslant 0$.  If one were
to use the parametrization of the energy density $\mathcal{E}$
through $f$ as suggested by Cohen \cite{Cohen:2005ea} the situation is
more complicated, as one would have to satisfy the nonlinear
differential constraint
\begin{equation}
  5ff''- 2f'^2\geqslant 0, \label{eq:Cohen:6}
\end{equation}
{\it cf.} Eq. (6) in Ref.~\cite{Cohen:2005ea}. 

Though not related by a strict Legendre transformation,
equations~(\ref{eq:dictionary}) show that the universal functions
$\g(\x)$ and $\h(\y)$ are similarly related, hence Fig.~\ref{fig:hg}
exhibits the same Maxwell construction properties.  Note that the
linear Maxwell construction only works with the functional form of
$\g(\x)$ as introduced in~(\ref{eq:E}): if one uses the function
$f(x)=\g^{5/3}(\x)$ introduced in~\cite{Cohen:2005ea}, then the
equivalent construction will involve nontrivial forms of $f(x)$ that
saturate the inequality (\ref{eq:Cohen:6})~\cite{Cohen:2005ea}:
\begin{equation}
f(x) = (A + Bx)^{5/3}.  
\end{equation}
For this reason, the function $\g(\x)$ is substantially simpler to
discuss than $f(x)$.  By further requiring saturation of
inequality~(\ref{eq:Cohen:6}) throughout the entire interval,
Cohen~\cite{Cohen:2005ea} obtained
\begin{equation}
  f(x) = \{1 + [(2\xi)^{3/5}-1]x\}^{5/3},
\end{equation}
which describes the case of a vanishing partially polarized region
PP$_{\a}$.

Finally, we discuss the physical significance of kinks in
$\mathcal{E}(n)$ which translate to flat regions of $\mathcal{P}(\mu)$
in the grand-canonical ensemble.  A kink in $\mathcal{E}(n)$ means
that, for a range of tangents (chemical potentials), the energy of the
ground state does not change.  In other words, the system is
insensitive to a range of chemical potentials.  For example, the
superfluid phase SF is insensitive to a range of chemical potential
splitting $\mu_{-}$ because of the physical gap in the spectrum.  The
flat regions in the grand canonical ensemble represents the same
physical state (same densities) that are stable over a range of
chemical potentials due to this gap.

As discussed in the main text, the system may or may not respond to
chemical potential differences strictly less than the gap $\mu_{-} <
\Delta$, but will definitely respond when $\mu_{-} > \Delta$.  This
gave us the bound $\Y_{1}$ on the transition parameter $\y_{1}$
(see~(\ref{eq:y1_constraint}) below).

\subsection{Maxwell Construction}
\begin{figure}[bp]
  \centering
  \psfrag{E}{$\mathcal{E}(n)$}
  \psfrag{n}{$n$}
  \ifprocessed{
    \includegraphics[width=\columnwidth]{figs_maxwell.eps}
  }{
    \includegraphics[width=\columnwidth]{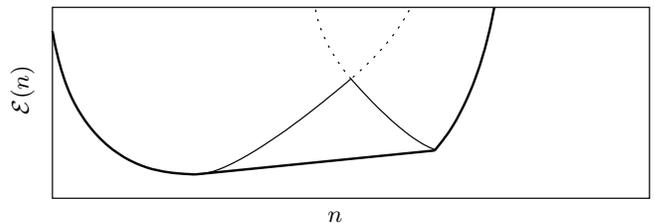}
  }
  \caption{\label{fig:maxwell} Maxwell construction for two competing
    phases in the microcanonical ensemble $\rho\in\{1,2\}$ with given
    energy densities $\mathcal{E}_{1}(n)$ and $\mathcal{E}_{2}(n)$.
    The right phase has a gap as signified by the kink.  The left
    phase has no gap.  A first-order phase transition connects the two
    phases.  Minimizing~(\ref{eq:Emin}) produces the thin solid curve,
    but this is not convex.  The Maxwell construction amounts to
    finding the convex hull---the thick solid line---which restores
    the convexity of $\mathcal{E}(n)$ required by the second law of
    thermodynamics.}
\end{figure}

As an example, let us consider the Maxwell construction for phase
coexistence in the microcanonical ensemble with a single species
``\a''.  The Maxwell construction for the case of two species at
unitarity is somewhat more complicated, but, as we discussed above,
involves the same type of linear construction if the function $\g(\x)$
is used.  This ensemble is constructed by minimizing the energy
density $\mathcal{E}(n)$ over all phases $\rho$ with fixed total
particle number $N=nV$, and volume $V$:
\begin{equation}
  \label{eq:Emin}
  \mathcal{E}(n) = \min_{\rho} \mathcal{E}_{\rho}(N/V). 
\end{equation}
In the microcanonical ensemble, this procedure is slightly
complicated by the possibility of phase coexistence.  Consider two
distinct pure phases $\rho\in\{1,2\}$ with given energy densities
$\mathcal{E}_{1}(n)$ and $\mathcal{E}_{2}(n)$: an example of two such
phases is shown in Fig.~\ref{fig:maxwell}.  Minimizing~(\ref{eq:Emin})
over these phases separately produces an energy density $\mathcal{E}(n)$ that is
not convex.  The Maxwell construction proceeds by constructing a
series of systems with fixed $N$ by combining a physical fraction $z$
of the system in phase $\rho_{1}$ with density $n_{1}$ and the
remaining fraction $1-z$ in phase $\rho_{2}$ with density $n_{2}$
subject to the constraint of fixed average density
\begin{equation}
  n = \frac{N}{V} = zn_{1} + (1-z)n_{2}.
\end{equation}
This leaves two degrees of freedom and defines a series of two-component mixed
phases that must also be considered in~(\ref{eq:Emin}) with energy density:
\begin{equation}
  \mathcal{E}(n) =
  z\mathcal{E}_{\rho_{1}}(n_{1})+(1-z)\mathcal{E}_{\rho_{2}}(n_{2}).
\end{equation}
Minimizing $\mathcal{E}(n)-\mu n$, where $\mu$ is a Lagrange
multiplier, leads to the linear segment shown in bold in
Fig.~\ref{fig:maxwell}, which is the convex hull of the energy
densities.  More specifically, by varying $n_{1,2}$ and $z$, (and
assuming that all derivatives exists), one obtains the slope
\begin{equation}
  \label{eq:Mqxwell_Equilibrium}
  \mu_c=\frac{\mathcal{E}_{\rho_{1}}(n_{1})-\mathcal{E}_{\rho_{2}}(n_{2})}{n_{1}-n_{2}}=
  \pdiff{\mathcal{E}_{\rho_{1}}(n_{1})}{n_{1}}=
  \pdiff{\mathcal{E}_{\rho_{2}}(n_{2})}{n_{2}}.
\end{equation}
This is simultaneously the slope of the line segment defining the
mixed phase, and the tangents to the two energy densities.  In the
case of the $\mathcal{E}_{\rho_{2}}$ shown in Fig.~\ref{fig:maxwell},
the function has a cusp at the point of transition.  In this case,
$\partial{\mathcal{E}_{\rho_{2}}(n_{2})}/\partial{n_{2}}$ should be
interpreted as the appropriate value in the interval defined by the
corresponding derivatives computed to the left and to the right of the
kink.  In~(\ref{eq:Mqxwell_Equilibrium})
$\partial\mathcal{E}_{\rho_{2}}/\partial{n_{2}}$ may assume any of
these values to satisfy the equilibrium condition.  This construction
guarantees that the potential $\mathcal{E}(n)$ will be convex as
required by the second law of thermodynamics.

The conditions of minimization are equivalent to the conditions that
the chemical potential and pressure of the two coexisting phases be
equal.  Note that phase coexistence occurs only along the linear
segment.  Along this entire segment, the tangent is the same.  Thus,
the entire region of phase coexistence is described by a single point
$\mu_{c}$ representing the phase transition in the grand-canonical
ensemble.  Furthermore, since the density changes discontinuously from
one side of the phase transition to the other, the pressure---though
continuous---will have a kink at $\mu_{c}$, consistent with a
first-order transition.

For this reason, phase structures are much simpler when considered in
the grand-canonical ensemble.  In this ensemble, the phase diagram
consists of only pure phases: all phase coexistence (mixtures) occur
along first-order phase transitions.

\subsection{Constraints}
Here we present a few more details about the extraction of the various
constraints in the main text.  We start with the bound $Y_{0}$.  Recall
that the transition is defined by chemical potential $\mu_{\b}$
required to keep out species ``\b''.  This is bounded by the energy
gained by adding the single particle $e_{0}$ which we estimate:
$\mu_{\b}\leq e_{0}$.  We must relate this to the chemical potential
$\mu_{\a}$ through the density of the free Fermi sea $n_{\a} = \beta
\mu_{a}^{3/2} \Rightarrow \mu_{a} = (n_{\a}/\beta)^{2/3} = \alpha
n_{\a}^{2/3}$.   We now
express the constraint $\mu_{\b}\leq e_{0}$ in terms of $\y_{0}$:
\begin{equation}
  \y_{0} = \frac{\mu_{\b}}{\mu_{\a}} \leq 
  \frac{e_{0}}{\alpha n_{\a}^{2/3}} = \Y_{0}.
\end{equation}
The bound $\Y_{1}$ follows from the constraint $\mu_{-} \leq \Delta$.
To express this in terms of the data, we need to express the chemical
potentials in terms of the normalization Fermi energy
$\varepsilon_{\text{F}} = \mu^{\text{FG}}_{+}$ of a free gas of the same density
$n_{+} = n_{\a}+n_{\b} = 2(2m\mu^{\text{FG}}_{+})^{3/2}/(6\pi^2)$:
\begin{equation}
  \varepsilon_{\text{F}} = \frac{1}{2m}\left[(3\pi^2)n_{+}\right]^{2/3}
\end{equation}
The relationship between the density of the symmetric phase SF and the
chemical potential $\mu_{+}$ is determined by the parameter $\xi$
from~(\ref{eq:PSF}):
\begin{equation}
  n_{+} = \pdiff{\mathcal{P}_{\text{SF}}(\mu_{+})}{\mu_{+}} = 
  \frac{2\beta}{\xi^{3/2}}\mu_{+}^{3/2}.
\end{equation}
Thus, we have simply $\varepsilon_{\text{F}}=\mu_{+}/\xi$.  We now
express the constraint:
\begin{equation}
  \frac{\mu_{-}}{\varepsilon_{\text{F}}} = 
  \xi\frac{\mu_{-}}{\mu_{+}} =
  \xi\frac{1-\y_{1}}{1+\y_{1}} \leq
  \frac{\Delta}{\varepsilon_{\text{F}}}.
\end{equation}
Solving for $\y_{1}$ gives the constraint
\begin{equation}
  \label{eq:y1_constraint}
  \y_{1} \geq \Y_{1} =
  \frac{\xi-\Delta/\varepsilon_{\text{F}}}
  {\xi+\Delta/\varepsilon_{\text{F}}}.
\end{equation}
These constraints transform directly into constraints on the
derivatives of $\g(\x)$ through the dictionary~(\ref{eq:dictionary}).
For example, when $\x=0$ we have
\begin{align}
  \y_{0} &\leq \Y_{0} &&\Rightarrow &
  \frac{g'(0)}{g(0)} = g'(0) &\leq \Y_{0},
\end{align}
where we have used the fact that $g(0) = 1$.  Likewise at $\x=1$ we
have
\begin{align}
  \y_{1} &\geq \Y_{1} &&\Rightarrow& 
  \frac{g'(1)}{g(1)+g'(1)} &\geq \Y_{1},
\end{align}
which gives a lower bound on $g'(1)$.  The upper bound on $g'(1)$ is
simply the condition that $\y_{1}\leq 1$ which we may impose by
symmetry:
\begin{align}
  \y_{1} &\leq 1 &&\Rightarrow&
  \frac{g'(1)}{g(1)+g'(1)} &\leq 1.
\end{align}
\subsection{Density Profiles}
\begin{figure}[b]
  \centering
  \psfrag{y0=y0}{$\y_{0}=\Y_{0}$}
  \psfrag{yc=y1}{$\y_{c}\approx\Y_{1}$}
  \psfrag{y1}{$\y_{1}$}
  \psfrag{y=mub/mua}{$\y=\mu_{\b}/\mu_{\a}$}
  \psfrag{hy}{$\h(\y)$}
  \psfrag{x=nb/na}{$\x=n_{\b}/n_{\a}$}
  \psfrag{gx}{$\g(\x)$}
  {\footnotesize
    \psfrag{0.8}{0.8}
    \psfrag{0.9}{0.9}
    \psfrag{1.0}{1.0}
    \psfrag{1.1}{1.1}
    \psfrag{1.2}{1.2}
    \psfrag{1.3}{1.3}
    \psfrag{-0.6}{-0.6}
    \psfrag{-0.4}{-0.4}
    \psfrag{-0.2}{-0.2}
    \psfrag{0.0}{0.0}
    \psfrag{0.2}{0.2}
    \psfrag{0.4}{0.4}
    \psfrag{0.6}{0.6}
    \psfrag{0.8}{0.8}
    \ifprocessed{
      \includegraphics[width=\columnwidth]{figs_gLobo.eps}
    }{
      \includegraphics[width=\columnwidth]{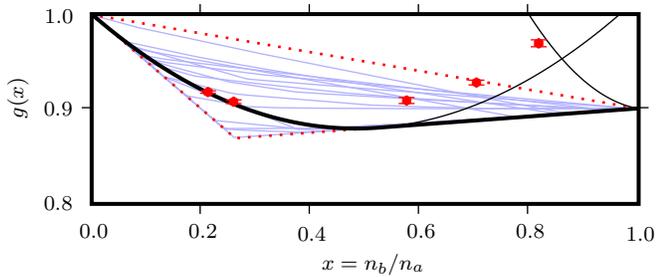}
    }
  }
  \caption{\label{fig:hgLobo} Monte Carlo variational upper bound on
    $\g(\x)$ from~\cite{LRGS:2006} plotted on top of the function $\g(\x)$
    from our Figure~\ref{fig:hg}.  Note the agreement from small
    polarizations indicating that our estimate for $\Y_{0}$ is
    consistent with their proper variational bound.  For larger
    polarizations, the true curve will lie below the results of
    Ref.~\cite{LRGS:2006} for two reasons: 1) The Maxwell construction for
    $g(x)$ (see Fig. (2) of~\cite{LRGS:2006}) and 2) The authors of
    Ref.~\cite{LRGS:2006} considered only normal Fermi partially polarized
    states.  As shown in~\cite{Bulgac:2006gh}, at $T=0$, partially
    polarized states will be superfluid.  This could noticeably lower
    the energy for substantial polarizations at unitarity.}
\end{figure}

We consider the following functional form for $\g(\x)$:
\begin{equation}
  \label{eq:gx}
  \g(\x) = 
  \begin{cases}
    g_{0} +g'_{0} x + ax^2 
    & \text{where } x \in [0,x_{T}],\\
    c + dx
    & \text{where } x \in [x_{T},1]
  \end{cases}
\end{equation}
The form of the function has been arbitrarily chosen between $0$ and
$x_{T}$ (for simplicity, we chose a quadratic polynomial) with the
correct intercept $\g(0)=1$ and slope $\g'(0)=\y_0$.  From this curve,
we proceed with the Maxwell construction by finding the line that
passes through $(x,\g(x))=(1,\g(1))$ and that is tangent to the given
curve $\g(x_{T})$ at the point $x_{T}$.  The resulting $g(x)$ is shown
with solid thick line in Fig.~\ref{fig:hg} and Fig.~\ref{fig:hgLobo}.

Once this curve is established, one can extract $\h(\y)$
using Eqs.~(\ref{eq:dictionary}).  The only complication arises when there
are kinks or linear segments.  In this case, there is a linear segment
of $\g(x)$ for $x\in[x_{T},1]$.  From~(\ref{eq:dictionary}) we see
that, throughout this region, $\y$ and $\h(\y)$ take on only a single
value: this is the first-order phase transition $y_1$.  This will
result in a kink in the function $\h(\y)$ at this location.

\begin{figure}[tbp]
  \centering
  \psfrag{Fully Paired (SF)}{Fully Paired (SF)}
  \psfrag{Vac}{Vac}
  \psfrag{Partially Polarized (PPa)}{Partially Polarized (PP$_{\a}$)}
  \psfrag{PPb}{PP$_{\b}$}
  \psfrag{Nb}{N$_{\b}$}
  \psfrag{Na}{N$_{\a}$}
  \psfrag{mua}{$\mu_{\a}$}
  \psfrag{mub}{$\mu_{\b}$}
  \psfrag{na,nb}{$n_{\a,\b}$}
  \psfrag{R}{$\tilde{R}$}
  {\footnotesize
    \psfrag{0}{0}
    \psfrag{0.0}{0.0}
    \psfrag{0.5}{0.5}
    \psfrag{1.0}{1}
    \psfrag{0.00}{0}
    \psfrag{0.25}{0.25}
    \psfrag{0.50}{0.50}
    \psfrag{R0}{$\tilde{R}_0$}
    \psfrag{R1}{$\tilde{R}_1$}
    \ifprocessed{
      \includegraphics[width=\columnwidth]{figs_profiles.eps}
    }{
      \includegraphics[width=\columnwidth]{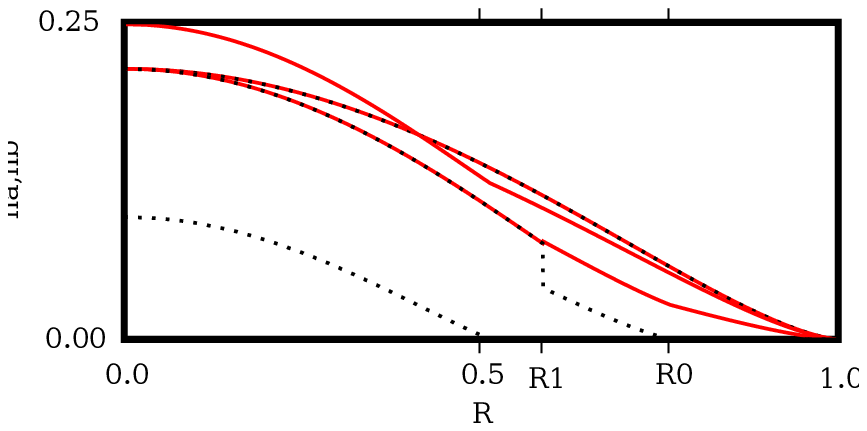}
      \includegraphics[width=\columnwidth]{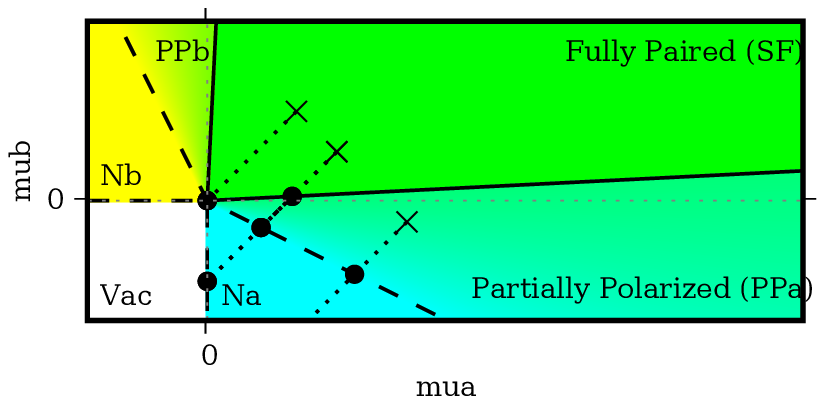}
    }
  }
  \caption{\label{fig:profiles} Density profiles $n_{a,b}$ of the two
    species in a spherical harmonic trap as a function radius
    $\tilde{R}=R/R_{\text{vac}}$ in units of the cloud radius
    $R_{\text{vac}}$ for thermodynamic function~(\ref{eq:gx}).  Density
    profiles are plotted for a fixed $\lambda_{+} =
    (\lambda_{\a}+\lambda_{\b})/2$ and a variety of chemical potential
    differences $\mu_{-} = \lambda_{-}$ ranging from $\lambda_{-}=0$
    (fully paired $n_a=n_b$ throughout the trap) to
    $\lambda_{-}=\lambda_{+}$ (no superfluid core).  The red solid
    lines are the majority species $n_{\a}$ while the black dotted
    lines are the minority species.  The critical radii for the
    intermediate profile have been denoted $\tilde{R}_{0,1}$.}
\end{figure}

To compute the density profiles in a trap, we first parametrize the
trap so as to establish the spatial variations of the chemical
potential.  Let us consider a spherical harmonic trap $V(R) \propto
R^2$ and consider the coordinate $\tilde{R}=R/R_{\text{vac}}$ where
$R_{\text{vac}}$ is the radius of the cloud.  The effective chemical
potentials are established by~(\ref{eq:mu_r}) once the Lagrange
multipliers $\lambda_{\a,\b}$ are chosen.  Once this is done, the
functional form of $\h(\y)$ can be used to directly map the position
$\tilde{R}$ to the densities using Eq.~(\ref{eq:densities}).  In
Fig.~\ref{fig:profiles} we show several trap density profiles with
the sample function~(\ref{eq:gx}).

\subsection{Extracting $g(x)$ from Experiment}
In principle, once the radial density profiles have be measured to
sufficient accuracy, one can extract the functional forms for $\g(\x)$
and $\h(\y)$.  For example, an experiment can extract the density
profiles $n_{\a,\b}(\vect{R})$ (see for example~\cite{SZSSK:2006}) for
a known trapping potential $V(\vect{R})$.  Using
relations~(\ref{eq:mu_r}) and~(\ref{eq:mub}), we have:
\begin{equation}
  \g^{2/3}[\x(\vect{R})]\g'[\x(\vect{R})] 
  = \frac{\mu_{\b}(\vect{R})}{\alpha n_{a}(\vect{R})^{2/3}} 
  = \frac{\lambda_{b}-V(\vect{R})}{\alpha n^{2/3}_{a}(\vect{R})}.
\end{equation}
This first-order differential equation may be integrated with the boundary
condition $\g(0)=1$ to find the function $\g(\x)$.  Finally,
$\lambda_{\b}$ will have to be fit to the trap profile.  Note, there are
many other ways to extract the same information: we have simply chosen
a simple method.  Other methods may be less sensitive to experimental errors for
example.  We leave it to future work to perform this extraction and
the accompanying error analysis. 
\subsection{Mean Field (Eagles-Leggett) Results}
In this section we consider the Eagles-Leggett mean-field
model~\cite{eagles}.  In this model, one can easily calculate $\xi$
and $\Delta$ (see for example~\cite{HS:2006}):
\begin{align}
  \label{eq:xi_MF}
  \xi_{MF} &= 0.5906, &
  \frac{\Delta_{MF}}{\varepsilon_{F}} &= 0.6864.
\end{align}
These determine the properties of the superfluid phase SF.  If we
consider only homogeneous and isotropic phases, then there are three
distinct phases: SF, N$_{\a,\b}$, and N$_{\a}$.\footnote{There is the
  possibility that other phases, like LOFF, compete with the partially
  polarized phase N$_{\a,\b}$. In the mean-field model it is unlikely
  that these phases will drastically alter the locations of the phase
  transitions.  The LOFF-like regions are always very thin, and we
  neglect these possibilities here.}  The SF phase is the usual
symmetric BCS-BEC crossover phase, the PP$_{\a}$ phase is a partially
polarized two-component Fermi liquid, and the N$_{\a}$ phase is a
fully polarized single-component Fermi liquid.  One of the
shortcomings of the mean-field crossover models is that they neglect
the Hartree-Fock contributions.

These terms enter the energy density as $\mathcal{E}_{HF} \approx gn_a
n_b$ where $g$ is the coupling constant.  In the limit of a
short-range interaction, one takes the range of the interaction $r_0
\rightarrow 0$ to zero while holding the vacuum inverse $s$-wave
scattering length $a^{-1}$ fixed.\footnote{The $a^{-1}=0$ limit at
  unitarity does not represent a singularity.}  This requires that the
interaction strength be taken to zero $g \sim \pi r_0/m \rightarrow
0$.  The densities $n_{\a,\b}$ contain no singularities and so the
Hartree-Fock contributions vanish.  These contributions may be
included in weak coupling by resumming particle-particle ladders to
obtain $\mathcal{E}_{HF} \approx 4\pi\hbar^2 a n_an_b/m$, but this
procedure may not be extrapolated to unitarity.

Thus, in mean-field, the partially polarized state simply has the
pressure of two independent free Fermi gases:
\begin{equation}
  \mathcal{P}_{\text{FG}_{2}}(\mu_{\a},\mu_{\b}) = 
  \tfrac{2}{5}\beta\left(\mu_{\a}^{5/2}+\mu_{\b}^{5/2}\right).
\end{equation}
Pressure equilibrium thus determines the phase transitions in this
model and we have the two conditions
\begin{subequations}
  \begin{align}
    \mathcal{P}_{\text{FG}_{2}}(\mu_{\a},\y_0^{MF}\mu_{\b}) &= 
    \mathcal{P}_{\text{FG}}(\mu_{\a}),\\
    \mathcal{P}_{\text{FG}_{2}}(\mu_{\a},\y_1^{MF}\mu_{\b}) &= 
    \mathcal{P}_{\text{SF}}(\mu_{+}).\label{eq:MF_SF/PP}
  \end{align}
\end{subequations}
These imply trivially that $\y^{MF}_0=0$.  Since the Hartree-Fock terms
vanish, there is no interaction energy and the phase transition
between N$_{\a}$ and PP$_{\a}$ occurs for $\mu_{\b}=0$.  The
SF/PP$_{\a}$ transition at $\y_1$ is governed by
condition~(\ref{eq:MF_SF/PP}).  Using Eq.~(\ref{eq:PSF}) and
$\mu_{\b}=y_0\mu_{\a}$, we have
\begin{equation}
   2\xi_{MF}^{-3/2}\left(\frac{1+\y^{MF}_{1}}{2}\right)^{5/2}=1+(\y^{MF}_1)^{5/2}.
\end{equation}
This may be solved numerically using Eq.~(\ref{eq:xi_MF}) to obtain
$\y^{MF}_{1}=0.1067$. To summarize: In the Eagles-Leggett mean-field
crossover model with homogeneous and isotropic phases, we have:
\begin{subequations}
  \begin{align}
    \y^{MF}_{0} &= \Y^{MF}_{0} = 0,&
    \y^{MF}_{c} &= 0.1051,\\
    \Y^{MF}_{1} &= -0.0750,&
    \y^{MF}_{1} &= 0.1067.
  \end{align}
\end{subequations}
Obviously the lower bound $\Y_1^{MF}$ is useless for our purposes.

\subsection{Calculation of $e_0$}
The Hamiltonian describing a system of $N_{\a}$ species ``\a'' fermions
interacting with one additional species ``\b'' fermion is:
\begin{equation}
  \op{H} = \op{H}_{\a}+\op{H}_{0} = 
  \left(\sum_{n=1}^{N_{\a}}\op{T}_{n}\right)+
  \left(\op{T}_0+\sum_{n=1}^{N_{\a}}\op{V}_{n,0}\right),
\end{equation} 
where $\op{T}_k$ are the corresponding kinetic energy operators and
$\op{V}_{n,0}$ is the potential between the fermion of species ``a''
with coordinate $\vect{r}_n$ with the fermion of species ``b'' with
coordinate $\vect{r}_0$.  Recall that we use the variational
wave function~(\ref{eq:wf})
\begin{equation}
  \Psi(\vect{r}_{0};\{\vect{r}_{n}\}) = 
  \Phi_{\text{SD}}(\{\vect{r}_{n}\})\phi(\vect{r}_{0};\{\vect{r}_{n}\}),
\end{equation}
where (\ref{eq:phi0})
\begin{equation}
  \phi(\vect{r}_{0};\{\vect{r}_{n}\}) = \sum_{n} A_{n}
  \frac{\exp{(-\kappa r_{0n})}}{r_{0n}}
\end{equation}
is an eigenfunction of the operator $\op{H}_{0}$:
\begin{equation}
  \op{H}_{0}\phi(\vect{r}_{0};\{\vect{r}_{n}\}) =
  -\frac{\hbar^2 \kappa^2(\{\vect{r}_{n}\})}{2m}
  \phi(\vect{r}_{0};\{\vect{r}_{n}\}).
\end{equation}

\begin{figure}[b]
  \centering
  \psfrag{dk/k (\%)}{$\delta\kappa/\kappa$ (\%)}
  \psfrag{Maximum displacement (units of lattice spacing)}
  {Maximum displacement (units of lattice spacing)}
  {\footnotesize
    \psfrag{0.0}{0.0}
    \psfrag{0.1}{0.1}
    \psfrag{0.2}{0.2}
    \psfrag{0.3}{0.3}
    \psfrag{0.4}{0.4}
    \psfrag{0}{0}
    \psfrag{1}{1}
    \psfrag{2}{2}
    \ifprocessed{
      \includegraphics[width=\columnwidth]{figs_kappa.eps}
    }{
      \includegraphics[width=\columnwidth]{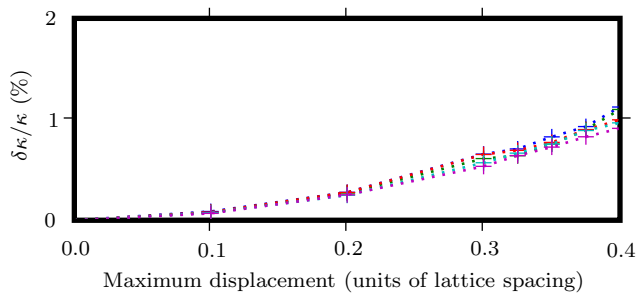}
    }
  }
  \caption{\label{fig:kappa}
    Relative change in $\kappa$ for randomly perturbed bcc lattices.
    The abscissa shows the maximum deviations each lattice site was
    displaced from equilibrium as a fraction of the lattice spacing.
    Curves are shown for cubic lattices from 10 to 14 
    sites per side.
    Each curve shows the maximum deviation over a sample of 30 random
    lattice configurations.
  }
\end{figure}
The Slater determinant is of course an eigenfunction of $\op{H}_{\a}$:
\begin{equation}
  \op{H}_{\a}\Phi_{\text{SD}}(\{\vect{r}_{n}\}) = 
  E_{FG}\Phi_{\text{SD}}(\{\vect{r}_{n}\}).
\end{equation}
Two additional terms arise from $\op{H}_{\a}$ acting on the product
wave function.  The first term is a cross-derivative which arises from
applying $\op{H}_{a}$ to both the Slater determinant and
$\phi(\vect{r}_0;\{\vect{r}_n\})$.  This gives a vanishing
contribution, since the integral over momenta
\begin{equation}
  \frac{\hbar^2}{m}\int_{k<k_F} 
  \frac{\d{\vect{k}}}{(2\pi)^3}
  \int \prod_{j=0}^{N_{\a}} \d{\vect{r}}_j
  \sum_{l=1}^{N_{\a}}(i\vect{k}\cdot \vect{\nabla}_l) 
  \phi^2(\vect{r}_0;\{\vect{r}_n\})
\end{equation}
is identically zero because of the symmetry of the Fermi sea.  The
second kind of term is due to $\op{H}_{\a}$ acting on
$\phi(\vect{r}_{0};\{\vect{r}_{n}\})$ and which pulls down derivatives
of $\kappa(\{\vect{r}_{n}\})$ and $A_n(\{\vect{r}_{n}\})$:
\begin{multline}
  \op{H}_{\a}\phi(\vect{r}_{0};\{\vect{r}_{n}\}) = 
  -\frac{\hbar^2\kappa^2(\{\vect{r}_{n}\})}{2m}
  \phi(\vect{r}_{0};\{\vect{r}_{n}\})\\
  +\text{ derivative corrections}.
\end{multline}
In what follows, we neglect the derivative corrections, but, as we
discuss below, we expect these to be small.  Within this
approximation, the expectation value of
this Hamiltonian is 
\begin{equation}
  \langle \Psi|\op{H}|\Psi\rangle  = E_{FG} +e_0 =E_{FG}
  -\frac{\hbar^2\kappa^2}{m}
\end{equation}
where $E_{FG}$ is the ground state energy of the system of $N_{\a}$
fermions alone.  This contribution arises by applying
$\sum_{n=1}^{N_{\a}}T_n$ to the Slater determinant alone.  It is
implied here that $\phi(\vect{r}_0;\{\vect{r}_n\})$ is normalized.

We have checked explicitly that at unitarity for several simple
lattice configurations (fcc, bcc) and for configurations in which the
positions of all the fermion species ``a'' were randomly changed from
the ideal lattice configurations within the Fermi hole, the values of
$\kappa$ varied by a few percent at most, see Fig. (\ref{fig:kappa}). 
This analysis thus confirms
our assumption that the fluctuations in a free Fermi gas of the number
density $n_a(\vect{r})$ do not affect in a noticeable manner the value
of $\kappa$.

For comparison, we replot the function $\g(\x)$ from Fig.~\ref{fig:hg}
and include the Monte Carlo data from~\cite{LRGS:2006}.  The authors of
Ref.~\cite{LRGS:2006} simulated a two-component polarized normal Fermi gas,
which provides a variational bound on the energy.  We thank
S. Giorgini for sending us their numerical results \cite{LRGS:2006}.

\begin{thebibliography}{10}

\bibitem{ZSSK:2005}
  M.W. Zwierlein, A.~Schirotzek, C.H. Schunck, and W.~Ketterle,
  \newblock Science {\bf 311}, 492 (2006),
  arXiv:cond-mat/0511197.

\bibitem{PLKLH:2005} 
  G.B. Partridge, W.~Li, R.~I. Kamar, Y.~an~Liao, and R.~G. Hulet,
  \newblock Science {\bf 311}, 503 (2006),
  arXiv:cond-mat/0511752.

\bibitem{ZSSK:2006} 
  M.W. Zwierlein, C.H. Schunck, A.~Schirotzek, and W.~Ketterle,
  \newblock Nature (London), \textbf{442}, 54 (2006),
  arXiv:cond-mat/0605258.

\bibitem{comments} M.W. Zwierlein and W.~Ketterle,
  \newblock Science, 314, 54 92006 (2006),
  arXiv:cond-mat/0603489;
  G.B. Partridge, W.~Li, R.~I. Kamar, Y.~an~Liao, and R.~G. Hulet,
  \newblock Science, 314, 54 (2006),
  arXiv:cond-mat/0605581.

\bibitem{meanfield} 
  P.F. Bedaque, H. Caldas, and G. Rupak,
  \newblock Phys.~Rev.~Lett.\ {\bf 91}, 247002 (2003),
  arXiv:cond-mat/0306694;
  T.~Mizushima, K.~Machida, and M.~Ichioka,
  \newblock \textit{ibid.}\ {\bf 94}, 060404 (2005),
  arXiv:cond-mat/0409417;
  J.~Kinnunen, L.M.~Jensen, and P.~T\"orm\"a,
  \newblock \textit{ibid.} {\bf 96}, 110403 (2006),
  arXiv:cond-mat/0512556;
  W.~Yi and L.-M. Duan,
  \newblock Phys.~Rev.~A {\bf 73}, 031604(R) (2006),
  arXiv:cond-mat/0601006;
  M.~Iskin and C.A.R. Sa~de~Melo, 
  \newblock Phys.~Rev.~Lett. {\bf 97}, 100404 (2006),
  arXiv:cond-mat/0604184;
  K.~Machida, T.~Mizushima, and M.~Ichioka,
  \newblock  \textit{ibid.}\ {\bf 97}, 120407 (2006),
  arXiv:cond-mat/0604339;
  L.M. Jensen, J.~Kinnunen, and P.~T\"orm\"a,
  \newblock 
  arXiv:cond-mat/0604424;
  C.-C. Chien, Q.~Chen, Y.~He, and K.~Levin,
  \newblock Phys.~Rev.~Lett.\ {\bf 97}, 090402 (2006),
  arXiv:cond-mat/0605039;
  W.-C. Su,
  \newblock 
  arXiv:cond-mat/0511183;
  L.~He, M.~Jin, and P.~Zhuang,
  \newblock Phys.~Rev.~B {\bf 73}, 214527 (2006),
  arXiv:cond-mat/0601147;
  X.-J. Liu and H.~Hu,
  \newblock Europhys. Lett. {\bf 75} 364 (2006),
  arXiv:cond-mat/0603011;
  C.-H. Pao and S.-K. Yip,
  \newblock J. Phys.: Condens. Matter {\bf 18} 5567 (2006),
  arXiv:cond-mat/0604530;
  M.~M. Parish, F.~M. Marchetti, A. Lamacraft, and B.~D. Simons, 
  \newblock Nature Physics 3, 124 (2007),
  arXiv:cond-mat/0605744.

\bibitem{Chevy:2006a}
  F.~Chevy,
  \newblock Phys.~Rev.~Lett. {\bf 96}, 130401 (2006),
  arXiv:cond-mat/0601122.

\bibitem{HS:2006}
  M.~Haque and H.T.C.~Stoof,  
  \newblock Phys. Rev. A {\bf 74}, 011602 (2006),
  arXiv:cond-mat/0601321.

\bibitem{Cohen:2005ea} 
  T.D.~Cohen,
  \newblock Phys.~Rev.~Lett. {\bf 95}, 120403 (2005),
  arXiv:cond-mat/0501080.
  
\bibitem{appendix} See the appendix or EPAPS Document No. for
  additional calculational details.  For more information on EPAPS,
  see http://www.aip.org/pubservs/epaps.html.

\bibitem{loff}
  P.~Fulde and R.~A. Ferrell,
  \newblock Phys.~Rev. {\bf 135}, A550 (1964).
  A.I.~Larkin and Y.N.~Ovchinnikov, 
  \newblock Sov.~Phys.~JETP {\bf 20}, 762 (1965).

\bibitem{Muther:2002mc}
  H.~M\"uther and A.~Sedrakian,
  \newblock Phys.~Rev.~Lett. {\bf 88}, 252503 (2002),
  arXiv:cond-mat/0202409.

\bibitem{Bulgac:2006gh}
  A.~Bulgac, M.M.~Forbes, and A.~Schwenk,
  \newblock Phys. Rev. Lett. {\bf 97}, 020402 (2006),
  arXiv:cond-mat/0602274.

\bibitem{Pandharipande}
  J.~Carlson, S.Y.~Chang, V.R.~Pandharipande, and K.E.~Schmidt,
  \newblock Phys.~Rev.~Lett. {\bf 91}, 050401 (2003),
  arXiv:physics/0303094;
  S.Y.~Chang, V.R.~Pandharipande, J.~Carlson, and K.E.~Schmidt,
  \newblock Phys.~Rev.~A {\bf 70}, 043602 (2004),
  arXiv:physics/0404115.

\bibitem{ABCG:2004}
  G.E.~Astrakharchik, J.~Boronat, J.~Casulleras, and S.~Giorgini,
  \newblock Phys.~Rev.~Lett. {\bf 93}, 200404 (2004),
  arXiv:cond-mat/0406113.

\bibitem{Carlson:2005kg}
  J.~Carlson and S.~Reddy,
  \newblock Phys.~Rev.~Lett. {\bf 95}, 060401 (2005),
  arXiv:cond-mat/0503256.

\bibitem{grad-lda} 
  T.N.\ De~Silva and E.J.\ Mueller,
  \newblock Phys.~Rev.~Lett. {\bf 97}, 070402 (2006),
  arXiv:cond-mat/0604638;
  A.~Imambekov, C.J. Bolech, M.~Lukin, and E.~Demler,
  \newblock Phys.~Rev.~A {\bf 74}, 053626 (2006),
  arXiv:cond-mat/0604423.

\bibitem{eagles} 
  D.R.~Eagles,
  \newblock Phys. Rev. {\bf 186}, 456 (1969);
  A.J.~Leggett, in {\it Modern Trends in the Theory of Condensed
    Matter}, edited by A. Pekalski and R. Przystawa,
  (Springer--Verlag, Berlin, 1980); 
  \newblock J. Phys. (Paris) Colloq. {\bf 41}, C7 (1980).

\bibitem{Chevy:2006b} F. Chevy, \newblock
  arXiv:cond-mat/0605751.

\bibitem{SZSSK:2006} Y. Shin, M.W.~Zwierlein, C.H.~Schunck,
  A.~Schirotzek, and W.~Ketterle, Phys. Rev. Lett. {\bf 97}, 030401
  (2006),
  arXiv:cond-mat/0606432.

\bibitem{LRGS:2006} C. Lobo, A. Recati, S. Giorgini and S. Stringari, 
  \newblock  Phys.~Rev.~Lett.\ {\bf 97}, 200403 (2006),
  arXiv:cond-mat/0607730.

\end{thebibliography}
\end{document}